\documentclass[12pt,a4paper]{iopart}
\usepackage{iopams,graphicx,bbm}

\newcommand{\Chi}{\mathrm{X}}
\newcommand{\assign}{:=}
\newcommand{\mathd}{\mathrm{d}}
\newcommand{\mathe}{\mathrm{e}}
\newcommand{\tmop}[1]{\mathrm{#1}}

\begin{document}
\title[Theory and experimental verification of Kapitza-Dirac-Talbot-Lau interferometry]{Theory and experimental verification of Kapitza-Dirac-Talbot-Lau interferometry}
\author{Klaus Hornberger$^{1,2}$, Stefan Gerlich$^{2}$,
Hendrik Ulbricht$^{2,3}$, Lucia Hackerm\"{u}ller$^{2,4}$, Stefan
Nimmrichter$^{2}$, Ilya V. Goldt$^5$, Olga Boltalina$^{5,6}$ and
Markus Arndt$^2$}
\address{$^1$ Arnold Sommerfeld Center for Theoretical Physics,
Ludwig-Maximilians-Universit\"{a}t M\"{u}nchen, Theresienstra{\ss}e 37,
80333 M\"{u}nchen, Germany}
\address{$^2$ Faculty of Physics, University of Vienna, Boltzmanngasse 5, A-1090 Wien, Austria}
\address{$^3$ School of Physics and Astronomy, University of Southampton, UK}
\address{$^4$ Johannes Gutenberg-Universit\"{a}t Mainz, Staudingerweg. 7, 55099 Mainz, Germany}
\address{$^5$ Department of Chemistry, Moscow State University, Moscow, 119992, Russia}
\address{$^6$ Department of Chemistry, Colorado State University, Fort Collins, CO, 80523, USA}
\ead{markus.arndt@univie.ac.at}

\date{\today}

\begin{abstract}
Kapitza-Dirac-Talbot-Lau interferometry (KDTLI) has recently been
established for demonstrating the quantum wave nature of large
molecules. A phase space treatment permits us to derive closed
equations for the near-field interference pattern, as well as for
the Moir{\'e}-type pattern that would arise if the molecules were
to be treated as classical particles. The model provides a simple
and elegant way to account for the molecular phase shifts related
to the optical dipole potential as well as for the incoherent
effect of photon absorption at the second grating. We present
experimental results for different molecular masses,
polarizabilities and absorption cross sections using fullerenes
and fluorofullerenes and discuss the alignment requirements. Our
results with C$_{60}$ and C$_{70}$, C$_{60}$F$_{36}$ and
C$_{60}$F$_{48}$ verify the theoretical description to a high
degree of precision.
\end{abstract}

\maketitle


\section{Introduction}
The quantum wave nature of matter has become a corner stone of
physics over many decades, and current interest in de Broglie
interferometry with
electrons~\cite{Tonomura1987a,Batelaan2007a,Sonnentag2007a},
neutrons~\cite{Rauch2000a}, atoms~\cite{Berman1997a,Cronin2008a},
and molecules~\cite{Arndt1999a,Arndt2009a} ranges from
demonstrating fundamental quantum phenomena to advanced
applications in the materials sciences and in quantum metrology.
All these experiments require optical elements for the coherent
manipulation of matter waves. While clean solid surfaces and bulk
crystal structures are well-adapted to the diffraction of
electrons and neutrons with de Broglie wavelengths in the range
of 1..1000\,pm, it is often necessary to tailor the beam
splitters, lenses, and wave guides to the specific particle
properties in atomic and molecular applications.

For complex molecules, nanofabricated gratings were demonstrated
to act as beam splitters for far-field
diffraction~\cite{Arndt1999a,Bruehl2004a} and near-field
interferometry~\cite{Brezger2002a}. However, these experiments
pointed already to the importance of van der Waals interactions
between the molecules and the diffraction grating, which largely
exceeds the effect observed with atoms~\cite{Grisenti1999a}
because of the high molecular polarizability and their
comparatively low velocity. The interaction time with a 500\,nm
thick grating amounts to only 5\,ns at a beam velocity of
100\,m/s, and yet the matter wave phase shift can attain the
value of several radians in the center of the slit opening. The
interaction effect gets even stronger close to the slit walls, to
a degree that the wave front distortion can no longer be
described by a phase shift alone~\cite{Nimmrichter2008a}. For
particles with increasing polarizability this strong influence of
the grating interaction leads to prohibitive requirements on the
velocity, as discussed in~\cite{Gerlich2007a}. It is therefore
appealing to replace material gratings by structures made of
light, which offer the additional advantage of being
indestructible, highly transparent, and easy to tune and modulate.

Bragg diffraction of {\em free electrons} at a standing light
field was already proposed by Kapitza and Dirac in
1933~\cite{Kapitza1933a}, but nearly seventy years passed before
the idea was experimentally implemented~\cite{Freimund2001a}. In
contrast to that, the first optical phase grating for atoms was
already realized in 1983~\cite{Moskowitz1983a,Gould1986a} when a
standing laser light field was tuned near to an atomic resonance
in order to perform Raman-Nath (`thin grating') diffraction of a
supersonic sodium beam. A related
investigation~\cite{Martin1988a} then focused on {\em atomic
diffraction} in the Bragg regime (`thick grating'). These ideas
were later extended to atom diffraction~\cite{Steane1995a} and
interferometry~\cite{Rasel1995a,Giltner1995a}, also to the time
domain~\cite{Cahn1997a,z1,z2}, and to the manipulation of {\em
Bose-Einstein} condensates~\cite{Ovchinnikov1999a,Deng1999a}.

The working principle of all phase grating examples is the same: A
coherent laser beam creates a periodic pattern of the electrical
field. This couples to the particle's polarizability, shifts the
energy and thus imprints a phase pattern on the transmitted matter
wave beam. Its evolution into a modulated particle density
distribution can then be observed further downstream.

Large, hot molecules in thermal beams often exhibit broad
absorption lines. Light will therefore mainly couple in a
non-resonant fashion. However, for most molecules one can still
find a suitable range of wavelengths where the light-molecule
coupling allows one to imprint a local matter wave phase shift of the
order of $\Delta \Phi=\pi$.

\begin{figure}
 \begin{center}
 \includegraphics[width=0.7\columnwidth]{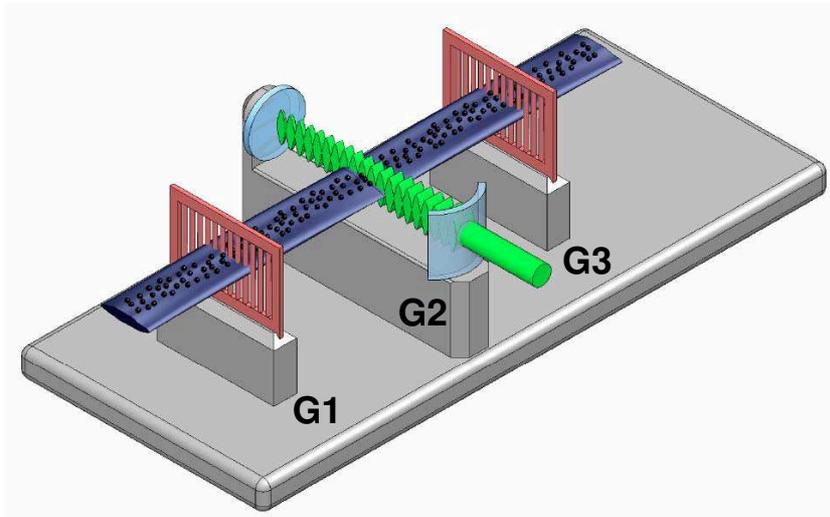}
  \caption{Setup of the Kapitza-Dirac-Talbot-Lau interferometer:
it consists of two material nanostructures (G1, G3) and a standing
 light wave (G2). The latter is realized by a cylindrical lens
 focusing a 532\,nm laser beam onto a mirror. The three structures
 (G1,G2,G3) have the same period of 266\,nm and are separated
 equidistantly by 105\,mm.  For  detection, the third grating (G3) is
 shifted in small steps over the molecular interference pattern.
 The transmitted molecules are detected in a quadrupole mass spectrometer.}\label{setup}
 \end{center}
 \end{figure}

The first application of optical phase gratings to {\em large, hot
molecules} was demonstrated with C$_{60}$ in a far-field
diffraction experiment~\cite{Nairz2001a}. The combination with
near-field diffraction was suggested in~\cite{Brezger2003a} and
recently implemented in a Kapitza-Dirac-Talbot-Lau interferometer
(KDTLI)~\cite{Gerlich2007a}, as shown in Fig.~\ref{setup}.

The general idea behind the KDTLI design has been described
elsewhere~\cite{Brezger2003a,Gerlich2007a}: A first absorptive
mechanical structure, G1, (in the present experiment $d=266\,$nm) prepares the required spatial coherence
for illuminating the optical phase grating, G2. Quantum
interference then explains the appearance of an approximate
self-image of G2 at the position of the third mask G3. This
molecular density pattern is scanned by shifting G3 across the
beam while counting the transmitted molecules as a function of
the mask position. This scheme was exploited to perform quantum
interference experiments up to extended polyatomic molecular
chains~\cite{Gerlich2007a} and to determine electrical molecular
properties~\cite{Hackermuller2007a,Gerlich2008a}.

In the present article we now provide a refined theoretical
description of Kapitza-Dirac-Talbot-Lau interference, putting
special emphasis on the proper incorporation of the influence of
photon absorption in the second grating. We give a closed
expression for both the quantum interference visibility and the
fringe contrast one would expect if a classical Moir\'e
description were correct. We then compare this to the measured
interference curves of C$_{60}$ and C$_{70}$ which are in nearly
perfect agreement with the quantum result. We also apply the KDTL
concept to studying the fluorofullerenes C$_{60}$F$_{36}$ and
C$_{60}$F$_{48}$. This allows us to assess the influence of mass,
absorption cross section, and optical polarizability on the
interference of large particles. In comparison to earlier
Talbot-Lau experiments~\cite{Hackermuller2003a}, the new
Kapitza-Dirac-Talbot-Lau interferometer now also allows us to
establish a significantly improved fringe contrast.

\section{Theory of the Kapitza-Dirac-Talbot-Lau interferometer}

The Kapitza-Dirac-Talbot-Lau interferometer is a derivative of
the standard Talbot-Lau
interferometer~\cite{Patorski1989a,Clauser1994a,Berman1997a,Brezger2002a},
obtained by substituting the central grating mask with the
optical phase grating created by a standing laser beam. In the
simplest configuration all gratings, material and optical, have
the same grating period $d$, given by one half of the laser
wavelength, $d=\lambda_L/2$. The passage of the matter wave beam
through each grating may thus transfer integer multiples of the
grating momentum $p_d=h/d= 2h/\lambda_L$ to the transverse motion
in the beam. These different diffraction orders interfere further
downstream, leading to a resonant enhancement at integer
multiples of the Talbot length $L_{\rm T}=d^2/\lambda_{\rm dB}$,
which is determined by the de Broglie wave length $\lambda_{\rm
dB}$ of the molecules (which ranges between 1\,pm and 5\,pm in our experiment). The emerging interference pattern at the
position of the third grating thus displays a strong dependence
of the interference fringe visibility on  $\lambda_{\rm dB}$, as
determined by the longitudinal velocity $v_z$ of the beam.

Unlike with material gratings, in the KDTLI we must also consider
the possibility that one or more laser photons are scattered
or absorbed while the molecule traverses the standing light wave.
The associated incoherent transfer of transverse momentum may
strongly blur the fringe pattern. If the absorption is followed
by an immediate isotropic reemission process, the transverse
momentum shift may take any value up to the photon momentum
$h/\lambda_L$. However, in many large molecules the absorbed
photon energy gets stored for a rather long time, either in
metastable excited states or, after rapid internal conversion, in
the vibrational degrees of freedom, which do not decay over the
time scale of the experiment. The associated transverse momentum
transfer is then an integer multiple of the photon momentum
$h/\lambda_L$, corresponding to one half of the grating momentum
$p_d$. An odd number of net photon momenta will thus kick the
molecular wave
such that the fringes get blurred maximally, while an even number
will have a much weaker effect. We note that related physics has
already been described in the context of far-field
diffraction~\cite{Pfau1994a} and Mach-Zehnder
interferometry~\cite{Chapman1995a} with atoms before.

In order to describe the interplay of coherent diffraction and the
incoherent effect of photon absorption we follow the phase space
formulation of Talbot-Lau interference based on the Wigner
function, as presented in {\cite{Hornberger2004a}}. It provides a
transparent representation of all relevant coherent and
incoherent phenomena, and it permits us to calculate the quantum
interference pattern on an equal footing with the possible
moir{\'e}-type structures, which might arise already due to
classical mechanics in this setup. This comparison with the
classical description is required if one wants to establish that
the observed fringe pattern in a molecule interference experiment
is caused by a genuine quantum interference effect.

\subsection{The light-grating interaction}\label{sec:lightgrating}

We start by collecting the necessary ingredients for describing
the effects of a light grating on the motion of a beam of
polarizable particles. Taking the direction of the particle beam
as the $z$-axis, we set the retro-reflected, basic Gaussian laser
mode in the perpendicular $x$-direction. The time averaged
intensity of the standing light wave is then given by
\begin{eqnarray}
  I \left( x, y, z \right) & = & \frac{8 P}{\pi w_y w_z} \exp \left( - \frac{2
  y^2}{w_y^2} - \frac{2 z^2}{w_z^2} \right) \sin^2 \left( \pi \frac{x}{d}
  \right),  \label{eq:intensity}
\end{eqnarray}
where  $w_y$ and $w_z $ denote the laser beam waists in the
vertical and the longitudinal direction, and $P$ is the laser
power. In the following, we assume the particle beam height to be
sufficiently small compared to $w_y$, such that the dependence on
the vertical $y$-direction can be safely neglected. This is approximately the case in our experiment, see below.

The standing light field will in general induce dispersive and
absorptive forces on a molecule. The first type, due to the
conservative optical dipole force, is described by the potential
\begin{eqnarray}
   V \left( x, z \right)  &
   = &  - \frac{2 \pi \alpha_{\omega}}{c}  I \left( x, 0,
  z \right) , \label{eq:Voptdip}
\end{eqnarray}
where $\alpha_{\omega}$ is the real part of the polarizability of
the particle at the laser frequency $\omega = 2 \pi c / \lambda_L$
(related to the polarizability in SI units by $
\alpha_{\tmop{SI}} \left( \omega \right) = 4 \pi \varepsilon_0
\alpha_{\omega}$).

Treating the effect of the grating potential in the eikonal
approximation, a traversing quantum wave acquires a position
dependent phase shift which is calculated by integrating the
potential along a straight line,

\begin{eqnarray}
  \phi \left( x \right) & = & - \frac{1}{\hbar} \int_{-\infty}^{\infty}  V
  \left( x, v_z t \right)\mathd t \hspace{0.3em} = \hspace{0.3em} \phi_0 \sin^2 \left(
  \pi \frac{x}{d} \right) .  \label{eq:phix}
\end{eqnarray}
The maximal shift
\begin{eqnarray}
  \phi_0 & = & 8\sqrt{2 \pi}\,  \frac{ \alpha_{\omega} }{\hbar c } \frac{P}{w_y v_z}
\label{eq:phi0def}
\end{eqnarray}
thus increases linearly with the optical polarizability and laser
power, and it is inversely proportional to the molecule velocity
$v_z$. The justification of this eikonal approximation and its
range of applicability are discussed in some detail in
\cite{Nimmrichter2008a}. As shown there, it is well justified for
the molecular masses and polarizabilities accessible with the
current experimental setup.

The  second type of momentum exchange between the light and the
molecules is the radiation pressure force due  to photon
absorption.
The molecules used in the present experiment are sufficiently
large and internally complex that it is justified to both ignore
any reemission and to take the absorption cross section constant
even after the absorption of several photons. In this case all
absorption events can be described as being independent and as
only determined by the absorption cross section
$\sigma_{\tmop{abs}}$ at the laser frequency $\omega=2\pi\nu$.
This cross section can often be related to the imaginary part of
the polarizability using Mie theory, $\sigma_{\tmop{abs}} = 4 \pi
\omega / c \times \tmop{Im} \left( \alpha \left( \omega \right)
\right)$, but we will treat $\sigma_{\tmop{abs}}$ as an
independent parameter in the following.

The photon absorption rate is determined by the incident photon
flux $I (x,z)/h\nu$ and $\sigma_{\tmop{abs}}$,
\begin{equation}
  \Gamma \left( x, z \right) =  \frac{\sigma_{\tmop{abs}}}{h \nu} I \left(  x,z \right).  \label{eq:Gamma}
\end{equation}
Below, the radiation pressure effect on the molecular beam will
be described by the position dependent mean number of absorbed
photons. The latter is obtained from the photon absorption rate,
in analogy to the eikonal approximation, by a straight
integration along the longitudinal motion of the molecule,

\begin{eqnarray}
  \bar{n} \left( x \right) & = & \int_{-\infty}^{\infty}  \Gamma \left( x, v_z t \right)\mathd t
  \hspace{0.3em} = \hspace{0.3em} n_0 \sin^2 \left( \pi \frac{x}{d}
  \right).
  \label{eq:nbarx}
\end{eqnarray}
The  maximum mean number of absorbed photons $ n_0$ is found in
the anti-nodes of the standing light wave, and it is given by
\begin{eqnarray}
  n_0 & = & \frac{8}{\sqrt{2 \pi}} \frac{ \sigma_{\tmop{abs}} \lambda_L}{hc}
  \frac{P}{w_y v_z } .  \label{eq:n0def}
\end{eqnarray}
The values of $\phi_0$ and $n_0$ defined in (\ref{eq:phi0def}) and
(\ref{eq:n0def}) are the two key parameters describing the
molecule-light interaction, and they will appear in the closed
formula for the quantum interference visibility below.

\subsubsection*{Classical description.}

We note that the momentum-changing effect of photon absorptions
does not differ in the quantum and the classical description of
the molecular motion. On the other hand,  the classical effect of
the optical dipole force due to the potential (\ref{eq:Voptdip})
should be treated in analogy  to the eikonal approximation of the
quantum case. This is done in terms of the momentum kick  $Q(x)$
obtained by integrating the dipole force along the same straight
line as in the eikonal treatment \cite{Hornberger2004a},
\begin{eqnarray}
  Q \left( x \right) & = & - \int_{-\infty}^{\infty} \frac{\partial V}{\partial x} \left( x, v_z
  t \right)  \mathd t \hspace{0.3em} = \hspace{0.3em} \frac{\pi \hbar}{d} \phi_0 \sin
  \left( 2 \pi \frac{x}{d} \right) ,  \label{eq:Qx}
\end{eqnarray}
with $\phi_0$ from (\ref{eq:phi0def}). (The  $\hbar$ in the
prefactor cancels Planck's constant in $\phi_0$ rendering the
equation classical.)

As we  will see below, the spatially periodic focusing of
classical particles due to the dipole potential \eref{eq:Qx} may
result in a regular molecule pattern behind the light grating,
though distinctly different from the quantum prediction. To
perform the quantum and the classical calculations it is
useful to formulate the effect of the grating passage in a
common framework, the Wigner-Weyl phase space representation.

\subsection{Phase space formulation of the light-grating transformation}

The most important part in describing the KDTL-interference is
the transformation of the particle beam state as it passes the
second grating. We consider the Wigner function
\begin{eqnarray}
  w\left( x, p \right) & = & \frac{1}{2\pi\hbar}\int \mathd s \,\rme^{2\pi i s p/\hbar}\langle x-\frac{s}{2}|\rho|x+\frac{s}{2}\rangle
\end{eqnarray}
of the transverse quantum state of motion $\rho$ of the molecular
beam \cite{Hillery1984a,Englert1994a,Hornberger2004a},
where $x$ and $p$ denote the position and momentum coordinates in phase space, and we
first assume the longitudinal velocity of the molecules to be
given by a definite value $v_z$.

After passing an arbitrary grating in eikonal approximation, the
transformed beam state can always be written as
\cite{Hornberger2004a}
\begin{eqnarray}
  w' \left( x, p \right) & = & \int \mathd p_0\, T \left( x, p - p_0 \right) w
  \left( x, p_0 \right)  .  \label{eq:generaltrafo}
\end{eqnarray}
That is, the momentum dependence of the quasi phase space
distribution gets modified by a convolution, while its position
dependence is at most affected by a multiplication.

Let us now discuss the grating transformation for an arbitrary,
$d$-periodic distribution of the light intensity. We first assume
that there is no absorption, $\sigma_{\rm abs} = 0$, so that the
grating transformation is entirely coherent. The phase shift
$\phi \left( x \right) $
then relates the wave function $\psi$ in front of the grating to
the one behind, $\psi' \left( x \right) = \exp \left( i \phi
\left( x \right) \right) \psi \left( x \right)$. In phase space
representation this coherent transformation is described by the
convolution kernel {\cite{Hornberger2004a}}
\begin{eqnarray}
  T_{\tmop{coh}} \left( x, p \right) & = & \frac{1}{2 \pi \hbar} \int \mathd
  s\,
  \mathe^{ips / \hbar}  \exp \left[ i \phi \left(  x - \frac{s}{2} \right)
- i \phi \left(  x + \frac{s}{2} \right) \right]   .
\label{eq:Tcohgen}
\end{eqnarray}
Noting the periodicity of $\phi(x)$ we define the Fourier
coefficients
\begin{eqnarray}
  b_j & = & \frac{1}{d}\int_{-d/2}^{d/2} \exp \left( i \phi \left( x \right) \right)
\rme^{- 2 \pi ij{x}/{d} }\rmd x , \label{eq:bjdef}
\end{eqnarray}
so that the coherent kernel takes the form
\begin{eqnarray}
  T_{\tmop{coh}} \left( x, p \right) & = & \sum_{j, m \in \mathbbm{Z}} b_j
  b_{j - m}^{\ast}  \exp \left( 2 \pi im \frac{x}{d} \right) \delta
  \left( p - \left( j - \frac{m}{2} \right) p_d \right) .  \label{eq:Tcohb}
\end{eqnarray}
This is a periodic comb of delta-functions separated by integer
multiples of the grating momentum $p_d=h/d$. It serves to
populate the different diffraction orders in
\eref{eq:generaltrafo} as the quantum wave passes the grating.

In a second limiting case,  we now consider the grating
transformation for a vanishing dipole force, $\alpha_\omega = 0$,
but maintain a finite absorption cross section.  It is reasonable
to assume that the final detection efficiency of the beam
particles is practically independent of the number of absorbed
photons. In this case, their motional state gets effectively
replaced by a statistical mixture whose components differ by
momentum translations of integer multiples of the photon momentum
$h / \lambda_L = p_d / 2$.  These multiples correspond to the
difference in the number of photons absorbed from the left and
from the right side. We denote as $\tmop{Prob} \left( k ; x
\right)$ the position-dependent probability distribution for the
exchange of $k \in \mathbbm{Z}$ net photon momenta. The mixture
can then be written in phase space representation as $w' \left(
x, p \right) = \sum_k \tmop{Prob} \left( k ; x \right) w \left(
x, p - kp_d / 2 \right) $. That is, the statistical
redistribution of the momenta due to photon absorption is
described by the incoherent kernel
\begin{eqnarray}
  T_{\tmop{abs}} \left( x, p \right) & = & \sum_{k \in \mathbbm{Z}}
  \tmop{Prob} \left( k ; x \right)  \delta \left( p - k \frac{p_d}{2}
  \right) .  \label{eq:Tinc1}
\end{eqnarray}
Expanding the periodic position dependence of $ \tmop{Prob}
\left( k ; x \right)$ in a Fourier series,
\begin{eqnarray}
  \tmop{Prob} \left( k ; x \right) & = & \sum_{j \in \mathbbm{Z}} P_j^{\left(
  k \right)} \mathe^{2 \pi ijx / d},  \label{eq:Probdef}
\end{eqnarray}
it takes the form
\begin{eqnarray}
  T_{\tmop{abs}} \left( x, p \right) & = & \sum_{k, j \in \mathbbm{Z}}
  \mathe^{2 \pi ijx / d} P_j^{\left( k \right)}   \delta
  \left( p - k \frac{p_d}{2} \right) .  \label{eq:Tinc2}
\end{eqnarray}
Let us now specify the probability distribution (\ref{eq:Probdef})
in terms of the mean number of absorbed photons $\bar{n} \left( x
\right)$. Since the absorption events are taken to be
statistically independent the probability $\tmop{Prob} \left( k ;
x \right)$ for the net gain of $k$ photon momenta is given by
\begin{eqnarray}
  \tmop{Prob} \left( k ; x \right) & = &
\sum_{n = 0}^{\infty} \tmop{Prob} \left( k|n \right) \frac{
\bar{n}^n (x)  }{n!}\, \rme^{ - \bar{n} (x)}   .  \label{eq:prob1}
\end{eqnarray}
Here, $\bar{n} \left( x \right)$ is the position-dependent mean
number of photon absorptions characterizing the Poissonian
distribution and $\tmop{Prob} \left( k|n \right)$ is the
probability for the net transfer of $k$ photon momenta towards
one side, conditioned on the absorption of exactly $n$ photons.
Since absorptions from the left and from the right occur with the
same probability in a standing light wave, the latter is given by
the distribution of a one-dimensional, balanced random walk with
$n$ steps,
\begin{eqnarray}
  \tmop{Prob} \left( k|n \right)  & = & \frac{1}{2^n}  \left\{
  \begin{array}{ll}
    {n \choose {(k + n)/2}}  & \mbox{if $n + k$ even}\\
    0 & \mbox{otherwise}.
  \end{array} \right.  \label{eq:prob2}
\end{eqnarray}
In order to perform the average over the Poissonian photon
distribution in \eref{eq:prob1} we first calculate the
characteristic function of (\ref{eq:prob2}) by means of the
binomial theorem,
\begin{eqnarray}
  \Chi \left( \xi |n \right) & = & \sum_{k \in \mathbbm{Z}} \mathe \tmop{xp}
  \left( - 2 \pi ik \xi \right)   \tmop{Prob} \left( k|n \right)
  \hspace{0.3em}  = \hspace{0.3em} \left[ \cos \left( 2 \pi \xi \right)
  \right]^n .  \label{eq:Chi1}
\end{eqnarray}
The characteristic function of the averaged distribution
(\ref{eq:prob1}) thus takes the simple form
\begin{eqnarray}
  \Chi \left( \xi ; x \right) & = & \sum_{n = 0}^{\infty} \frac{\left[ \bar{n}
  \left( x \right) \cos \left( 2 \pi \xi \right) \right]^n}{n!} \exp \left( -
  \bar{n} \left( x \right) \right) \hspace{0.3em} \nonumber\\
  & = & \hspace{0.3em} \exp \left\{
  - \bar{n} \left( x \right) \left[ 1 - \cos \left( 2 \pi \xi \right) \right]
  \right\}.  \label{eq:Chi2}
\end{eqnarray}
The inverse Fourier transform of (\ref{eq:Chi2}) yields the
required probability (\ref{eq:prob1}) in terms of the mean number
of absorbed photons,
\begin{eqnarray}
  \tmop{Prob} \left( k ; x \right) & = & \exp \left( - \bar{n} \left( x
  \right) \right) I_k \left( \bar{n} \left( x \right) \right),
  \label{eq:Probkx}
\end{eqnarray}
where the $I_n \left( x \right)$ are modified Bessel functions of
the first kind.

So far, the integral kernels $T_{\tmop{coh}}$ and
$T_{\tmop{abs}}$, which describe the coherent and the incoherent
part of the light grating interaction, were discussed separately,
see   Eq.~(\ref{eq:Tcohb}) and (\ref{eq:Tinc2}). For realistic
molecules the dispersive and the dissipative light forces
coexist, and their combined contribution is described in the
eikonal approximation by a single transformation
(\ref{eq:generaltrafo}), whose kernel is given by the convolution
of $T_{\tmop{coh}}$ and $T_{\tmop{abs}}$,
\begin{equation}
T \left( x, p \right) = \int \mathd q\,T_{\tmop{coh}} \left( x, p
- q \right) T_{\tmop{abs}} \left( x, q \right) .
\label{eq:ConvAbsDip}
\end{equation}

\subsubsection*{Classical description.}

As an advantage of the phase space formulation, it is easy to
describe in the same framework how the molecules would move if
they were classical particles. One merely replaces the Wigner
function by the classical phase space distribution, which is a
proper probability density. Both the quantum Wigner function and
the classical phase space distribution experience the same
shearing transformation as they evolve freely between the optical
elements, see the discussion in \cite{Hornberger2004a}. Also the
passage through a grating can be expressed in the same form
\eref{eq:generaltrafo} in both cases, though the integral kernels
differ of course. The classical kernel due to the dipole force
takes the form (\ref{eq:generaltrafo})
\begin{eqnarray}
  T_{\tmop{cl}} \left( x, p \right) & = & \delta \left( p - Q \left( x \right)
  \right),  \label{eq:Tcl}
\end{eqnarray}
where $Q(x)$ is the classical momentum kick of Eq.~(\ref{eq:Qx}).
The effect of a  photon absorption, on the other hand, is
described by the same kernel (\ref{eq:Tinc1}) as in the quantum
case, since it effects the same momentum change on the motional
state, irrespective whether the center-of-mass motion is
described by classical or quantum dynamics. To obtain the
combined effect of dispersive and absorptive light forces, one
can again concatenate the two transformations.

\subsection{Evaluating the Kapitza-Dirac-Talbot-Lau effect}
We are now in the position to calculate the interference pattern
expected for the KDTLI in the same spirit as it was done for
purely coherent grating interactions in {\cite{Hornberger2004a}}.
The final beam state is obtained by applying the appropriate
sequence of free evolution and grating transformations to its
Wigner function. Starting with a spatially completely incoherent
but monochromatic beam in front of the first grating one thus
obtains the spatial density distribution in front of the third
grating by a final integration over the momentum variable,
\begin{eqnarray}
  w_3 \left( x \right) & \propto & \sum_{k \in \mathbbm{Z}} \int \mathd x_0 \mathd p \,
   T_1( x_0) T_{\tmop{coh}} \left( x -
  \frac{p}{p_z} L, 2 p - \frac{x - x_0}{L} p_z - k \frac{p_d}{2} \right) \nonumber\\
  &  & \times\tmop{Prob}\left( k, x - \frac{p}{p_z} L \right) .  \label{eq:wxgen}
\end{eqnarray}
Here $L$ is the distance between the gratings and  $T_1(
x_0)\in\{0,1\}$ denotes the binary function which specifies the
transmission of the first material grating. The latter serves to
imprint a density modulation onto the beam, thus creating the
required spatial coherence downstream at the light grating. We
characterize the material grating mask by the Fourier coefficients
\begin{eqnarray}
A_j & := & \frac{1}{d}\int_{-d/2}^{d/2} T_1( x) \rme^{- 2 \pi
ij{x}/{d} }\rmd x,
  \label{eq:Adef}
\end{eqnarray}
where $d$ is the common period of all gratings.

In order to evaluate the interference pattern \eref{eq:wxgen} it
is now convenient to define the coherent {\em Talbot-Lau
coefficients} in terms of the Fourier coefficients
\eref{eq:bjdef} which describe the phase shift due to the optical
dipole potential,
\begin{eqnarray}
  B_m \left( \xi \right) & \assign & \sum_{j \in \mathbbm{Z}} b_j b_{j -
  m}^{\ast}   \exp \left( - 2 \pi i \left[ j - \frac{m}{2} \right] \xi
  \right) .  \label{eq:Bmdef}
\end{eqnarray}
The incoherent effect of the light grating is best accounted for
through the characteristic coefficients associated to the
Fourier coefficients of the periodic probability distribution
(\ref{eq:Probdef}). They are given by
\begin{eqnarray}
  \chi_j \left( \xi \right) & \assign & \sum_{k \in \mathbbm{Z} } P_j^{\left(
  k \right)} \exp \left( - 2 \pi i k \xi \right) ,  \label{eq:chidef}
\end{eqnarray}
and they serve to define the general Talbot-Lau coefficients,
which include the effect of absorption,
\begin{eqnarray}
  \hat{B}_m \left( \xi \right) & := & \sum_{n \in \mathbbm{Z}}  B_n \left( \xi \right)
  \chi_{m - n} \left( \frac{1}{2} \xi \right) .  \label{eq:Bhatdef}
\end{eqnarray}
The factor $\frac{1}{2}$ in (\ref{eq:Bhatdef}) reflects the fact a
single photon has a momentum that equals one half of the grating
momentum $p_d$. As one expects, the convolution
\eref{eq:Bhatdef}  reduces to the coherent expression
\eref{eq:Bmdef} if absorption can be neglected, i.e., for $\chi_m
\left( \xi \right) = \delta_{m, 0}$.

Inserting the Fourier expressions \eref{eq:Tcohb} and
\eref{eq:Probdef} into \eref{eq:wxgen} the integrations can now
be carried out by retaining the resonant contributions. This
yields the  interference pattern in terms of the coefficients
(\ref{eq:Adef}) and (\ref{eq:Bhatdef}),
\begin{eqnarray}
  w_3 \left( x \right) & = & \sum_{\ell \in \mathbbm{Z}}
  A^{\ast}_{\ell}    \hat{B}_{2 \ell} \left( \ell
  \frac{L}{L_{\rm T}} \right) \exp \left( 2 \pi i \ell \frac{x}{d} \right)
  .  \label{eq:w3qm}
\end{eqnarray}
Here $L_{\rm T}=d^2/\lambda_{\rm db}$ denotes the \emph{Talbot
length}, which gives the characteristic length scale for
near-field interference.

One records the beam intensity behind the third grating as a
function of the lateral position $x_s$, $S \left( x_s \right)
\propto \int \mathd x T_3 \left( x - x_s \right) w_3 \left( x
\right)$. Since the first and third grating are identical in our
experiments, $ T_1 \left( x \right) =T_3 \left( x \right)$,  the
expected interference signal reads
\begin{eqnarray}
  S \left( x_s \right) & = & \sum_{\ell \in \mathbbm{Z}}  \left(
  A^{\ast}_{\ell}  \right)^2  \hat{B}_{2 \ell} \left( \ell
  \frac{L}{L_{\rm T}} \right) \exp \left( 2 \pi i \ell \frac{x_s}{d} \right)
  .  \label{eq:Sqm}
\end{eqnarray}

\subsubsection*{Classical description.}
Using the same general formalism as above the classical result is
obtained by replacing the Wigner function by the classical phase
space density and the kernel $T_{\tmop{coh}} \left( x, p \right)$
in (\ref{eq:wxgen}) by its classical counterpart (\ref{eq:Tcl}).
The evaluation of the corresponding moir{\'e}-type density
distribution suggests to introduce the {\em classical
coefficients}
\begin{eqnarray}
  C_m \left( \xi \right) & = & \frac{1}{d} \int_{- d / 2}^{d / 2} \mathd x
   \exp \left( - 2 \pi im \frac{x}{d} \right) \exp \left( - 2 \pi i
  \frac{Q \left( x \right)}{p_d} \xi \right) .  \label{eq:Cmdef}
\end{eqnarray}
They are the classical analogue of the Talbot-Lau coefficients
\eref{eq:Bmdef}, but clearly lacking an interference phase factor.
Performing the same steps as above, the classical prediction for
the signal behind the third grating thus assumes a form analogous
to (\ref{eq:Sqm}),
\begin{eqnarray}
  S_{\tmop{cl}} \left( x_s \right) & = & \sum_{\ell \in \mathbbm{Z}} \left(
  A_{\ell}^{\ast} \right)^2  \hat{C}_{2 \ell} \left( \ell
  \frac{L}{L_{\rm T}} \right) \exp \left( 2 \pi i \ell \frac{x_s}{d}
  \right),  \label{eq:Scl}
\end{eqnarray}
where $L_{\rm T}=d^2 m v_z/h$, with $m$ the  molecular mass and
$v_z$ their longitudinal velocity. Note that Planck's constant
appearing in $L_{\rm T}$ cancels against the one from $p_d=h/d$
showing up  in \eref{eq:Cmdef}; it is kept here to maintain the
close analogy with the quantum result. Like in
(\ref{eq:Bhatdef}), the possibility of photon absorption is
accounted for in \eref{eq:Scl} by a convolution with the
characteristic coefficients (\ref{eq:chidef}),
\begin{eqnarray}
  \hat{C}_m (\xi) & = & \sum_n
  C_n \left( \xi \right) \chi_{m - n} \left( \frac{1}{2} \xi \right) .
  \label{eq:Chatdef}
\end{eqnarray}

\subsection{Closed expressions for the sinusoidal light grating}

The results obtained so far are valid for gratings with arbitrary
eikonal phase shifts and momentum kick distributions. We now
focus on the complex light grating of our experiment, as defined
by the intensity distribution \eref{eq:intensity}. Their special
form will yield closed formulas for the Talbot-Lau coefficients
(\ref{eq:Bhatdef}) and their classical analogues
\eref{eq:Chatdef}.

The sinusoidal $x$-dependence of the phase shift (\ref{eq:phix})
implies that the Fourier coefficients are determined by the
integer Bessel functions,
\begin{eqnarray}
  b_m & = &  \left( - i \right)^m  \mathe^{i \phi_0 / 2} J_m \left(
  \frac{\phi_0}{2} \right) .  \label{eq:bm}
\end{eqnarray}
The summation for the coherent Talbot-Lau coefficient
(\ref{eq:Bmdef}) can be carried out by means of Graf's addition
theorem for Bessel functions {\cite{Erdelyi1954a}}. This leads to
\begin{eqnarray}
  B_m \left( \xi \right) & = & J_m \left( - \phi_0 \sin \left( \pi \xi \right)
  \right),  \label{eq:Bexp}
\end{eqnarray}
indicating that all the Talbot-Lau coefficients are real.

It is instructive to compare this to the corresponding
coefficients (\ref{eq:Cmdef}) of the classical formulation. It
follows immediately from (\ref{eq:Qx}) that they are given by
\begin{eqnarray}
  C_m \left( \xi \right) & = & J_m \left( - \pi \phi_0 \xi \right) .
  \label{eq:Cexp}
\end{eqnarray}
Comparing the the quantum expression \eref{eq:Bexp} and the
classical one \eref{eq:Cexp}, we see that both coefficients
assume the same limiting form if the Talbot parameter $\xi$ is
much smaller than unity. They do however strongly deviate for
$\xi \geqslant 1$. The distinguishing quality of the {\em quantum
wave} coefficients (\ref{eq:Bexp}) is their periodicity in $\xi$,
which gives rise to the characteristic Talbot-Lau recurrences.
{\em Classical} particles show no such recurrences since their
coefficients (\ref{eq:Cexp}) exhibit no periodicity in $\xi$.

We move on to evaluate the characteristic coefficients
(\ref{eq:chidef}). There is no obvious way to express the Fourier
coefficients $P_n^k$ from Eq.~(\ref{eq:Probdef}) in closed form.
However, the coefficients $\chi_m \left( \xi \right)$ can be
expressed as the Fourier transform of the characteristic function
(\ref{eq:Chi2}) with respect to position,
\begin{eqnarray}
  \chi_m \left( \xi \right) & = & \int_{- 1 / 2}^{1 / 2} \mathd \tau \exp
  \left( - 2 \pi im \tau \right) \Chi \left( \xi, \tau d \right) .
  \label{eq:chiexp}
\end{eqnarray}
Due to the sinusoidal position dependence of the mean photon
number (\ref{eq:nbarx}) the integration can be carried out,
yielding a modified Bessel function,
\begin{eqnarray}
  \chi_m \left( \xi \right) & = & \exp \left( - n_0 \sin^2 \left( \pi \xi
  \right) \right) I_m \left( n_0 \sin^2 \left( \pi \xi \right) \right)\,.
\end{eqnarray}

The Talbot-Lau coefficients in the presence of absorption can now
be obtained by performing the summation in (\ref{eq:Bhatdef}).
This can be done using an addition theorem for mixtures of
regular and modified Bessel functions, which can be derived from
Graf's addition theorem. It reads, for $u, v \in \mathbbm{R}, u
\neq v$,
\begin{eqnarray}
  \left( \frac{v - u}{v + u} \right)^{n / 2} J_n \left( - \tmop{sgn} \left( u
  + v \right) \sqrt{v^2 - u^2} \right) & = & \sum_{k \in \mathbbm{Z}} I_{k +
  n} \left( u \right) J_k \left( v \right)   \label{eq:theorem}
\end{eqnarray}
and yields a real number also for $\left| v \right| < \left| u
\right|$ since $J_n \left( iu \right) = i^n I_n \left( u \right)$.

Using (\ref{eq:Bexp}) and (\ref{eq:chiexp}), and noting $J_{- n}
\left( z \right) = J_n \left( - z \right)$, we thus obtain the
general coefficients of the Kapitza-Dirac Talbot-Lau
interferometer, which incorporate the effect of photon absorption.
They are given by
\begin{eqnarray}
  \hat{B}_m \left( \xi \right) & = & \exp \left( - \zeta_{\tmop{abs}}(\xi) \right)
  \left( \frac{\zeta_{\tmop{coh}}(\xi) - \zeta_{\tmop{abs}}(\xi)}{\zeta_{\tmop{coh}}(\xi) +
  \zeta_{\tmop{abs}}(\xi)} \right)^{m / 2} \nonumber\\
  &  & \times J_m \left( - \tmop{sgn} \left[
  \zeta_{\tmop{abs}}(\xi) + \zeta_{\tmop{coh}}(\xi) \right] \sqrt{\zeta_{\tmop{coh}}^2(\xi) -
  \zeta_{\tmop{abs}}^2(\xi)} \right)  \label{eq:Bhatexp}
\,.
\end{eqnarray}
Here, the  coherent diffraction effect of the dipole force is
described by the function
\begin{eqnarray}
  \zeta_{\tmop{coh}}(\xi) & = & \phi_0 \sin \left( \pi \xi \right)
  \label{eq:zetacoh}
\end{eqnarray}
 and the incoherent effect of absorption is accounted for by
\begin{eqnarray}
  \zeta_{\tmop{abs}}(\xi) & = & n_0 \sin^2 \left( \frac{\pi}{2} \xi \right) \,.
  \label{eq:zetaabs}
\end{eqnarray}

\subsubsection*{Classical description.}
The coefficients for the classical motion (\ref{eq:Chatdef}) can
be obtained the same way starting from (\ref{eq:Cexp}) and
(\ref{eq:chiexp}). Given the relation between the classical and
the quantum coefficients (\ref{eq:Cexp}) and (\ref{eq:Bexp}), it
is not surprising that the $\hat{C}_m \left( \xi \right)$ assume
a similar form as the $\hat{B}_m \left( \xi \right)$ in
(\ref{eq:Bhatexp}). The only difference is that
$\zeta_{\tmop{coh}}(\xi)$ is replaced by
\begin{eqnarray}
  \zeta_{\tmop{cl}}(\xi) & = & \phi_0 \pi \xi
\, ,  \label{eq:zetacl}
\end{eqnarray}
which lacks the periodicity in the Talbot parameter $\xi$ shown by
(\ref{eq:zetacoh}).

\subsection{Discussion of the theoretical results }

\subsubsection{The fringe visibility}

Using the above results it is now easy to calculate the expected
quantum interference pattern \eref{eq:Sqm} and the corresponding
classical prediction \eref{eq:Scl}. However, for the parmeters of
our experiment the patterns are well described by a sine curve so
that it is sufficient to characterize the experimentaly observed
pattern by the contrast of a sinusoidal fit. This sinusoidal
fringe visibility  can be calculated as the ratio of the first
two Fourier coefficients of the fringe pattern, $\mathcal{V} = 2 |
{S_1}/{S_0}|$.

We denote by $f$ the open fraction  (i.e., the ratio between the
single slit width and the grating period) of the first and the
third grating, so that the grating coefficients \eref{eq:Adef}
are given by $A_{\ell} = f\,\tmop{sinc} \left( \ell \pi f
\right)$. The quantum fringe visibility then takes the form
\begin{eqnarray}
  \mathcal{V}_{\rm qm} & = &
   2\; \tmop{sinc}^2( \pi f) \left| \hat{B}_2 \left(
  \frac{L}{L_{\rm T}} \right) \right|  \label{eq:Vqm}
\end{eqnarray}
with the coefficient $\hat{B}_2$  given by~(\ref{eq:Bhatexp}).

Similarly, the fringe pattern expected form classically moving
molecules has a visibility $ \mathcal{V}_{\rm
cl}=2\,\tmop{sinc}^2(\pi f)|\hat{C}_2 ({L}/{L_{\rm T}})|$ which
is obtained from (\ref{eq:Vqm}), if we replace the function of
coherent diffraction  $\zeta_{\tmop{coh}}(\xi)$ by the classical
function $\zeta_{\tmop{cl}}(\xi)$  defined in
Eq.~(\ref{eq:zetacl}).

\subsubsection{Comparing the classical and quantum predictions.}

\begin{figure}[tbh]
   \includegraphics[width=0.7\paperwidth]{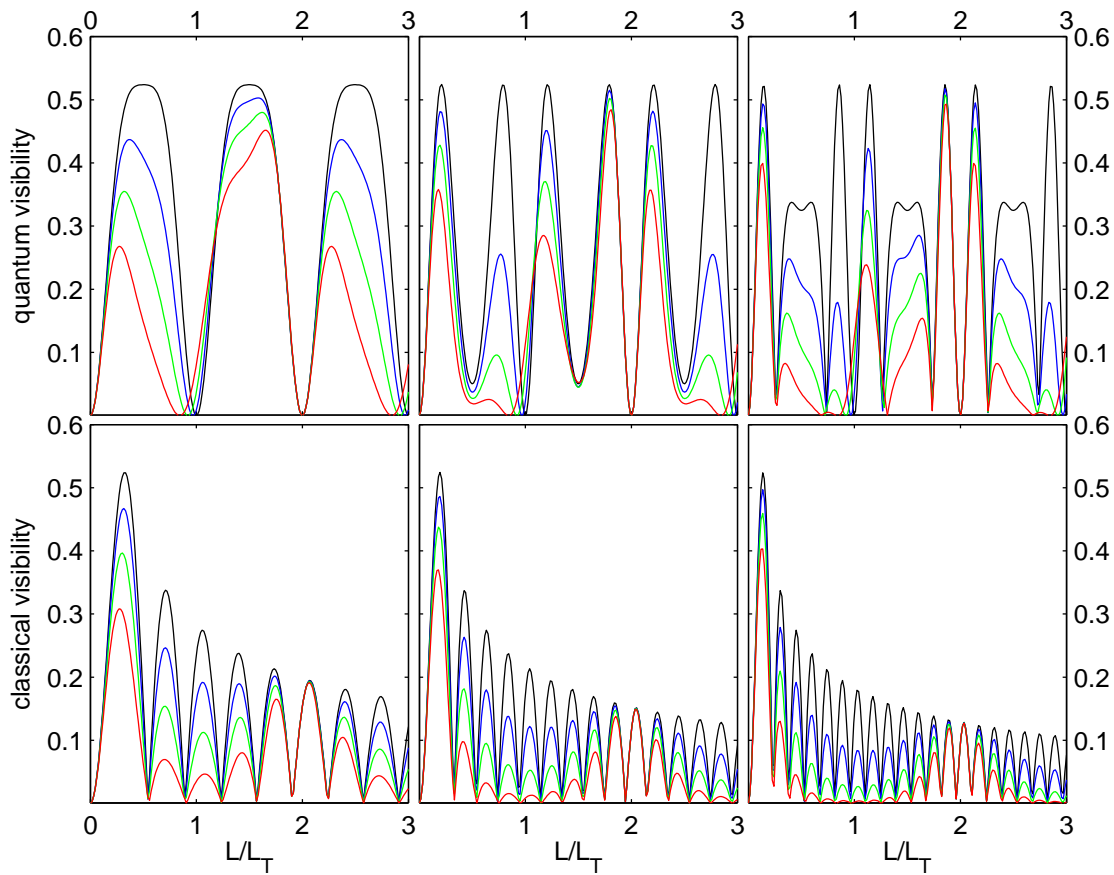}
  \caption{Fringe visibilities as obtained from the quantum (top)
  and classical (bottom) description of the molecular motion in
a KDTLI. The abscissa value $L / L_{\rm T}$ is proportional to the
molecular de Broglie wave length; it scales as $v_z^{-1}$, also in
the classical case. The material gratings are chosen to have an open
fraction of $f = 0.42$~\cite{Gerlich2007a}. The maximal optical
phase shift increases from $\phi_0 = 3$ (left panel), $\phi_0 = 5$
(middle) to $\phi_0 = 7$ (right). Each line in each panel describes
the fringe visibility for a specific maximal mean number of absorbed
photons $n_0$, which is given, from the top to the bottom lines, by
$n_0/\phi_0= 0\%, 10\%, 25\%,\mathrm{and}\,50\%$. The quantum and
the classical fringe visibilities differ markedly both in their
value and in their functional dependence, except for the limit of
high velocities, i.e. small de Broglie wave lengths $L / L_{\rm
T}\to 0$. The visibility peaks are strongly affected by photon
absorption.} \label{fig:theorie1}
\end{figure}

Let us now see how the quantum interference pattern differs from
the fringe pattern expected from classically evolving particles.
Figure \ref{fig:theorie1} compares the corresponding visibilities
$ \mathcal{V}_{\rm qm}$ and $ \mathcal{V}_{\rm cl}$ as one varies
the de Broglie wave length or velocity. The latter is specified
by $L/L_{\rm T}=L/d^2\times\lambda_{\rm dB}$ both in the quantum
and the classical case. For the material gratings we assume an
open fraction of $f = 0.42$ , like in our experiment.

The most important feature of the quantum result (top row in
Fig.~\ref{fig:theorie1}) is that the visibilities are generally
much greater than the classical calculation (bottom row). They
also display more structure if the strength of the dipol
potential increases (left column to right column), a consequence
of the intricate near field interference process. At first sight,
it may seem surprising that a fringe visibility would be observed
at all  in this setup if the molecules were moving as classical
particles. This is due to a moir{\'e}-type effect, where the
light grating acts as a periodic structure of lenses  focusing
the classical trajctories. Note that these classical visbilities
are systematically suppressed in the `quantum regime' $L / L_{\rm
T}\gg 1$. They coincide with the quantum result only in the
`classical limit' $L / L_{\rm T}\to 0$ of a vanishing de Broglie
wave length.

One also observes in Fig.~\ref{fig:theorie1} that the visibility
peaks are affected rather differently by the possibility of
photon absorption. Close to even multiples of $L / L_{\rm T}$ the
classical contrast remains essentially unaffected by absorption,
while the quantum visibility vanishes identically at all integer
multiples of $L / L_{\rm T}$ in the absence of absorption. It is
a curious result of our theory that a certain fringe pattern can
be obsered even if the dipole force can be neglected compared to
photon absorption, $\phi_0 \rightarrow 0$. The quantum and the
classical predictions coincide in this case, and one expects a
sinusoidal visibility given by
\begin{eqnarray}
  \mathcal{V}_{\tmop{abs}} & = & 2 \,\tmop{sinc}^2 \left( \pi f \right) \exp
  \left[ - \zeta_{\tmop{abs}}\left(\frac{L}{L_{\rm T}}\right) \right] I_2 \left[ \zeta_{\tmop{abs}}\left(\frac{L}{L_{\rm T}}\right) \right] .
\end{eqnarray}
It is greatest if the maximum mean number of absorbed photons
equals $n_0 = 4.65$. The visibility then amounts to
$\max(\mathcal{V}_{\tmop{abs}})=0.236 \times \tmop{sinc}^2 \left(
\pi f \right)$, i.e., a value of 24\% cannot be exceeded by this
effect.

\begin{figure}[tbh]
\includegraphics[width=0.7\paperwidth]{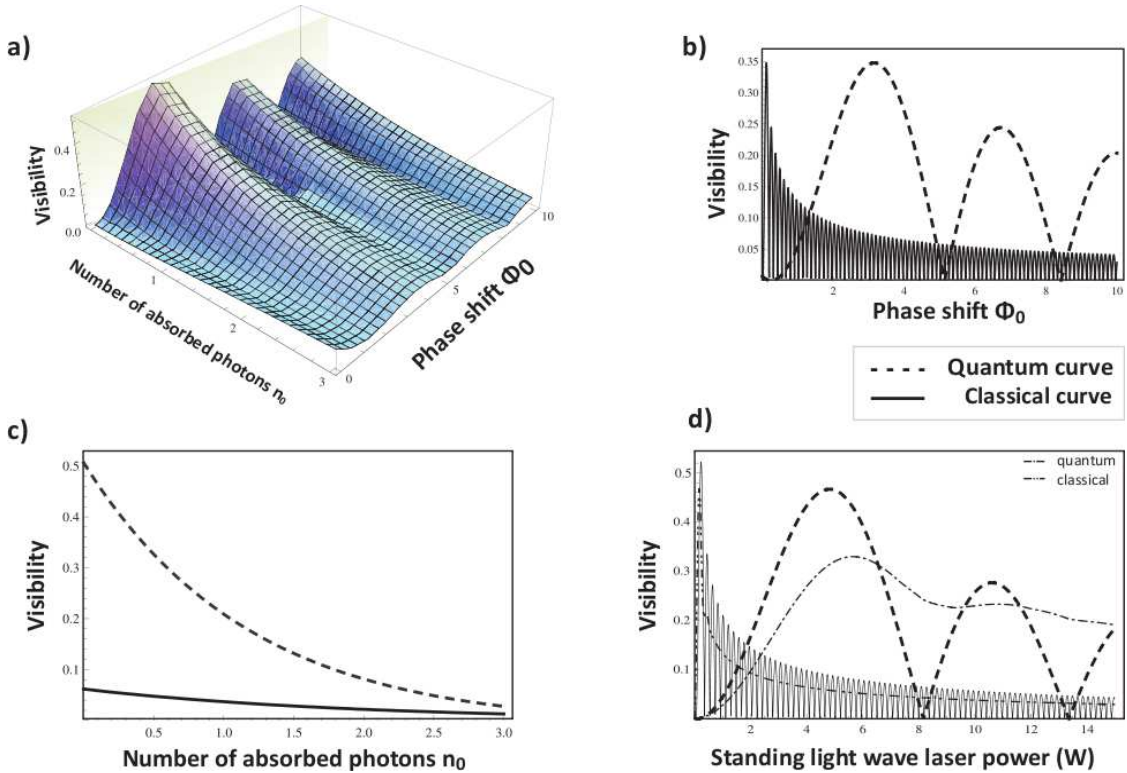}
\caption{Interference visibility for a fixed value of $L/L_{\rm
T}=8.5$ and $f=0.42$ for C$_{60}$, which corresponds to a velocity
of 97~m/s. Part a) shows the quantum interference visibility as a
function of $\Phi_{0}$ and $n_{0}$. Panels b) - d) display cuts
through this figure, along with the smaller classical fringe
visibility (solid lines). b) is the cut along $n_{0}$~=~0.5, c)
shows the cut for fixed $\Phi_{0}=2.7$, and d) is a cut along the
surface shown in a), which corresponds to a linear increase in
the power of the light grating according to equations
(\ref{eq:phi0def}) and (\ref{eq:n0def}). Also shown in d) is the
effect of a velocity distribution $\Delta\,v/v=10\%$ on the
quantum (dashed dotted curve) and classical (dash double dotted
curve) visibility.}\label{fig:theorie2}
\end{figure}

Finally, Fig.~\ref{fig:theorie2} shows for a fixed value of $L /
  L_{\rm T}$ how the visibilities depend on the molecular
properties, which are summarized in the dipole force phase
$\phi_0$ and the absorption number $n_0$. We choose $L / L_{\rm
T}$= 8.5, which corresponds to C$_{60}$ fullerenes at a velocity
of 97\,m/s. One observes that at fixed velocity the parameter
dependence is less complicated than the wave length dependence of
Fig.~\ref{fig:theorie1}. Since the molecular velocity is easy to
control this implies that KDTL-Interference can be used to measure
the molecular polarizability and the absorption cross section by
varying the intensity of the light grating
\cite{Gerlich2007a,Hackermuller2007a,Gerlich2008a}.

So far, the molecular beam was assumed to be characterized by a fixed longitudinal velocity $v_z$. The case of a finite velocity spread is easily incorporated by averaging the interference patterns of the monochromatic theory with the measured velocity distribution in the beam. This also applies to the sinusoidal visibilities, since the zeroth Fourier component of the interference pattern is independent of the velocity.

\section{Verification of the model using fullerenes and fluorofullerenes}
\begin{figure}[tbh]
{\begin{center}
 \includegraphics[width=0.7\columnwidth]{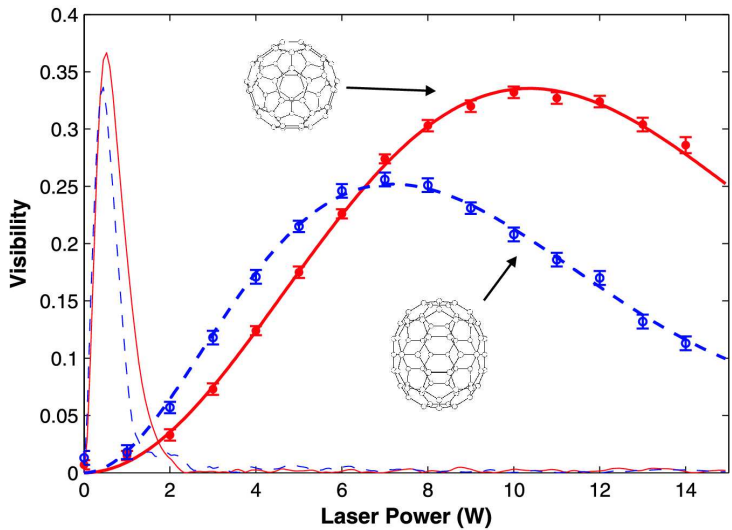}
  \caption{Power dependence of the interference contrast for the
fullerenes C$_{60}$ (filled circles) and C$_{70}$ (open circles).
The points represent the weighted average of three consecutive
measurements, the error bars depict one standard deviation. Bold
lines display the quantum expectations, obtained by weighting
Eq.\,\eref{eq:Vqm} with the experimentally observed velocity
distribution and fitting $\alpha$ and $\sigma_0$. The
corresponding classical expectations are shown as thin lines. The
solid lines identify the theoretical predictions for C$_{60}$
while the dashed lines correspond to C$_{70}$.}\label{fig:exp1}
\end{center}}
\end{figure}

In a first experiment a mixture of the fullerenes C$_{60}$ (720
amu) and C$_{70}$ (840 amu) was co-sublimated in a thermal source
(see Fig.~\ref{setup}) at a temperature of 910~K. By alternating
the setting of the detecting quadrupole mass spectrometer between
the masses of the two molecules, the interference patterns of
either species were recorded, one after the other, before the
laser power was shifted to the next point in the series. This
way, we assured that the standing light wave conditions were the
same for both molecules. The result is shown in
Figure~\ref{fig:exp1}. Each data point represents the weighted
average of three consecutive measurements, where the confidence
intervals of the individual sine fits serve as weights. The
single patterns were recorded over four full sine periods with
ten points per period and two seconds integration time per point.
The mean count rates per second amounted to 740 for C$_{60}$ and
505 for C$_{70}$.

The velocity of the particles was selected by limiting their path
to the associated free flight parabola in the Earth's
gravitational field~\cite{Arndt2001a}. This yields an approximately Gaussian 
velocity distribution. The mean velocities in the
experiment were determined to be 202~m/s for C$_{60}$ and 194~m/s
for C$_{70}$ with velocity spreads of 27\% and 25\%, respectively
($\Delta v/v$, FWHM).
The interferometer setup is characterized by a grating period of $d$=266\,nm (corresponding to a laser wave length of 532\,nm), 
a grating separation of $L$=105\,mm, a molecular beam height of about 200\,$\mu$m, and a beam width of about 1\,mm.
The slit widths are assumed to be $85$\,nm in G1 and $110$\,nm in G3.  

Both visibility curves reproduce the quantum expectations
accurately (bold lines, Eq.~(\ref{eq:Sqm})), while being in
distinctive disagreement with the classical prediction (thin
lines, Eq.~(\ref{eq:Scl})). We emphasize that the present result
signifies a noticeable improvement over previous
measurements~\cite{Gerlich2007a}. We attribute the enhanced
interference contrast mainly to further improvement of the highly
critical adjustment of the machine as outlined in
Section~\ref{alignmentsection}.

From the perfect accordance with the theory we also deduce that
decoherence due to collisions with particles of the background
gas is negligible in our current experiments. Following
\cite{Hornberger2003a} we estimate an effective cross section for
collisions of $\sigma_{\rm eff} = 4.2\times10^{-17}~\rm m^{2}$ for
C$_{70}$. The experiment is conducted under pressures below
$10^{-8}\,\mathrm{mbar}$, which results in a mean free path of
more than 17~meters. This corresponds to a reduction of the
effective visibility of ${\cal V}_{\rm eff}>0.98~{\cal V}_{0}$. Since
the effective collisional cross section is mainly governed by the
polarizability of the molecule rather than its geometrical size,
an even smaller reduction of the visibility can be expected for
the other species discussed in this article.

We also observe, and again in good agreement with our model, that
the fringe contrast of the more absorptive C$_{70}$ decays
significantly more rapidly than the contrast of C$_{60}$ when we
increase the laser power. The increasing number of absorbed
photons fills in the interference minima with shifted
interference curves, thus washing out the accumulated
interference pattern.

It may come as a surprise that, in spite of the higher absorption
cross section, C$_{70}$ exhibits actually a higher interference
contrast than C$_{60}$ at lower laser powers. This can be
explained by the optical polarizability which, according to our
present measurement, amounts to $\alpha_{\rm AC}=114{\rm \AA}^3$
for C$_{70}$ and is thus 31\% higher than for C$_{60}$. This
results in a larger phase shift in the optical grating and thus
leads to a shift of the entire curve to the left in
Fig.~\ref{fig:exp1}.

\begin{figure}[tbh]
{\begin{center}
\includegraphics[width=0.7\columnwidth]{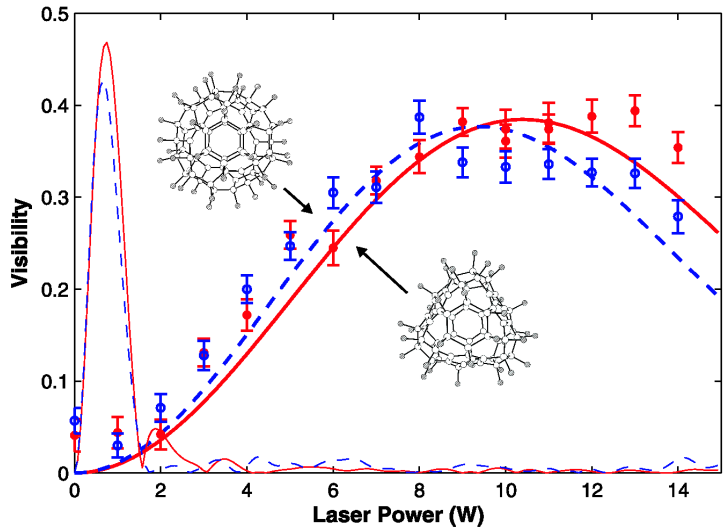}
\caption{Power dependence of the interference contrast for the
fluorofullerenes C$_{60}$F$_{36}$ (filled circles) and
C$_{60}$F$_{48}$ (open circles). The circles represent the
weighted average of three consecutive measurements, the error
bars depict one standard deviation of the shot noise. Bold lines
display the quantum expectations. The values used for the
parameters $\alpha$ and $\sigma_0$ correspond to a best fit and
are depicted in Tab.~\ref{CompareMolecules}. Classical
expectations are shown as thin lines. Solid lines correspond to
C$_{60}$F$_{36}$, dashed lines to
C$_{60}$F$_{48}$.}\label{fig:exp2}
\end{center}}
\end{figure}

The experiment was repeated in two additional and separate runs with
the fluorofullerenes C$_{60}$F$_{36}$ and C$_{60}$F$_{48}$. The
C$_{60}$F$_{36}$ sample was synthesized by the method described
in~\cite{Boltalina1996a} with a compositional purity of $>95\%$ as
determined by mass spectrum analysis. It contains two major isomers
of C$_3$ and C$_1$ symmetry in the approximate ratio 2:1, as well as
one minor isomer of T symmetry (ca. 5\%)~\cite{Popov2006a}. The
three isomers of C$_{60}$F$_{36}$ show very similar
polarizabilities, ranging between $62 \rm\AA$ and $65 \rm\AA$,
according to calculations performed with
Gaussian~\cite{Gaussian2003a}. Depending on the specific isomer,
C$_{60}$F$_{36}$ may posses a dipole moment of up 1.2 Debye. The
sample of C$_{60}$F$_{48}$ was purchased from Prof. L. Sidorov,
Moscow. The synthesis and characterization was done according
to~\cite{Boltalina2000a}. C$_{60}$F$_{48}$ is formed predominantly
as a single isomer of D$_3$ symmetry, with the minor isomer of S$_6$
symmetry comprising ca. 5\%~\cite{Popov2006a}. For both isomers of
C$_{60}$F$_{48}$ the simulations yield virtually identical
polarizabilities and no dipole moment. Both fluorofullerene samples
were produced before the year 2003 but remained intact compounds
over this period, as proven by mass spectra.

It is noteworthy that an earlier experiment in a pure Talbot-Lau
configuration succeeded already with C$_{60}$F$_{48}$ but at limited
contrast~\cite{Hackermuller2003a}. The present setup however
substantially outperforms its predecessor: The more sophisticated
KDTLI scheme, an improved count rate, and a better vibration
insulation with respect to the earlier experiment allowed for the
first time to achieve the full expected quantum contrast for both
C$_{60}$F$_{36}$ and C$_{60}$F$_{48}$, as shown in
Fig.~\ref{fig:exp2}. The larger error bars with respect to the
C$_{60}$-C$_{70}$ measurement are mainly a consequence of the lower
count rates of only 60 and 75 per second for C$_{60}$F$_{36}$ and
C$_{60}$F$_{48}$, respectively. The temperature was kept a 590~K for
both molecules.

The recorded velocities for C$_{60}$F$_{36}$ and C$_{60}$F$_{48}$
were 130~m/s and 116~m/s with velocity spreads of 16\% and 18\%,
respectively. Although smaller velocities tend to make the
experiment more susceptible to vibrations, drifts and
misalignment, no significant drop of the measured visibility
below the theoretical expectation was observed.

\begin{table}[bth]
\begin{center}
\begin{tabular}{|p{2.7cm}|p{2.7cm}|p{2.7cm}|p{2.7cm}|p{2.7cm}|}
\hline
Molecule & C$_{60}$ & C$_{70}$ & C$_{60}$F$_{36}$ & C$_{60}$F$_{48} $\\
\hline
$\sigma_{\mathrm{abs}} [10^{-22}\mathrm{m}^2]$& $2.8\pm0.3\pm0.3$ & $24.9\pm1.1\pm2.7$ & $<0.6$ & $<0.5$ \\
\hline
$\alpha_{\rm opt} [\AA^3]$ (Exp.)& $87.1\pm0.5\pm9.7$ & $114.2\pm0.9\pm12.7$ & $60.3\pm1.0\pm6.7$ & $60.1\pm0.8\pm6.7$ \\
\hline
$\alpha_{\rm stat} [\AA^3]$ (Lit.)& $88.9\pm0.9\pm5.1$~\cite{Berninger2007a} & $108.5\pm2.0\pm6.2$~\cite{Berninger2007a} & 62-65~\cite{Gaussian2003a} & 63~\cite{Gaussian2003a} \\
\hline
$\alpha [\rm \AA^3]$ (Lit.)& $89.2$~\cite{Guha1996a} & $109.2$~\cite{Guha1996a} & -- & -- \\
\hline
$\alpha [\rm \AA^3]$ (Lit.)& $90$~\cite{Eklund1995a} & $118.4$~\cite{Eklund1995a} & -- & -- \\
\hline
$\alpha [\rm \AA^3]$ (Lit.)& $98.2$~\cite{Sohmen1992a} & $122.6$~\cite{Sohmen1992a} & -- & -- \\
\hline
\end{tabular}
\caption{Molecular parameters as derived from the best fit of the
theory including statistical and systematic errors. The data are
in very good agreement with the values provided by literature,
where published, or with simulations performed with
Gaussian~\cite{Gaussian2003a}. The simulation yields slightly
different values values for the three conformers of
C$_{60}$F$_{36}$, ranging between $62\rm\AA^3$ and $65\rm\AA^3$.
Note that the published and calculated values represent {\em
static} polarizabilities while our experiment yields the {\em
optical} polarizability at the laser wavelength of 532\,nm.}
\end{center} \label{CompareMolecules}
\end{table}

In Tab.~\ref{CompareMolecules} we compare the optical properties
of all four particles that were extracted from a best fit of the
quantum curves to the experimental data. All values are in good
agreement with the parameters determined in independent
experiments ~\cite{Berninger2007a,Hackermuller2007a} and with
molecular simulations carried out using Gaussian. The remarkably
small statistical errors indicate that our method offers the
capacity for high precision metrology experiments with heavy
molecules. The accuracy is, however, currently limited by the
systematic errors which are primarily governed by the accuracy of
the measurement of the power ($\pm~5\%$) and the waist
($\pm~10\%$) of the diffracting laser beam.

The decrease in polarizability from C$_{60}$ to C$_{60}$F$_{48}$
is in good accordance with the observation that fluorinated
molecules in general show a reduced polarizability-to-mass ratio
and correspondingly lower inter-molecular binding, lower
sublimation enthalpies and higher vapor pressures at a given
temperature~\cite{Boltalina1999a}. It is also important to see
that the fluorine shell reduces the absorption cross section at
the wavelength of the diffracting laser beam to a negligible
value. Our measurement thus allows us to extract information
about the effect of fluorination on the electronic properties of
fullerenes.


\section{Alignment requirements for precision
experiments}\label{alignmentsection} Matter wave interferometry with
large molecules operates with de Broglie wavelengths in the range of
a few picometers and grating periods as small as a few hundred
nanometers. As a result of that, the interferometer alignment has to
be considered carefully. The following section is therefore devoted
to a short assessment of the constraints on the experimental
precision.

\subsection{Equality of grating periods}
If the first two gratings have only slightly different lattice
periods, the interference pattern spacing will not match the
period of the third mask and the contrast will be reduced. A
period mismatch as small as one per mille leads already to half a
fringe shift between slit one and five hundred.

In practice, all grating periods must be, on average, equal to
better than 0.05\,nm, i.e about the diameter of a hydrogen atom.
This condition enters both the choice of the grating manufacturing
process and the alignment of the yaw angle for all gratings.
Modern photo-lithography and etching procedures allow to reach
this level of precision. The gratings for our experiments were
produced by Dr. Tim Savas at MIT and `nm$^2$' Inc, Cambdridge,
Massachusetts, and independently checked by Ibsen photonics,
Denmark. The gratings were fabricated to be 0.3\,nm wider than
the period of the standing light wave in order to allow for later
yaw adjustments.

\subsection{Transverse grating shifts and grating roll angles}
The lateral position $\Delta x_{\rm i}$ of all gratings relative
to each other determines the final location of the fringe pattern.
For Talbot-Lau interferometry we can define a phase of the
near-sinusoidal interferogram which is determined by the relative
shift of the molecular density maxima with regard to the openings
of the third grating. For geometrical reasons this phase is
determined by (e.g.~\cite{Oberthaler1996a})
\begin{equation}
\phi= k_{\rm d} (\Delta x_{1}\, - 2 \Delta x_{2}\,+\, \Delta
x_{3})
\end{equation}
with $k_{\rm d}=2\pi/d$. In a symmetrical setup such as ours, with
$L_1 = L_2$, the prefactor of the first and third grating must be
equal while the second grating's shift enters twice.
\begin{figure}
\begin{center}
\includegraphics[height=0.95\textwidth]{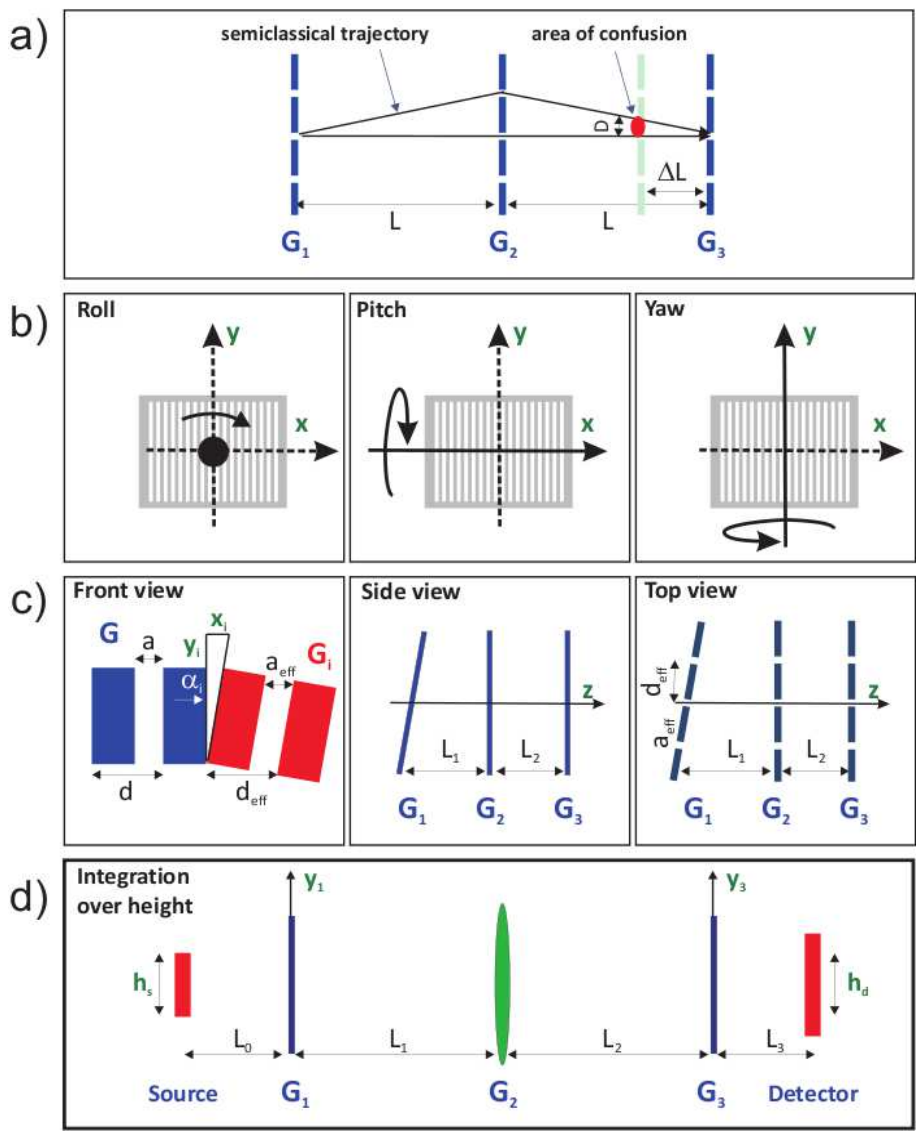}
\end{center}
\caption{Illustration of the alignment considerations: a) An
imbalance of grating separations leads to blurred interferences;
b) and c) Illustration of the grating's motion during roll, pitch
and yaw. d) Distances relevant for the assessment of grating roll:
molecules can pass the gratings at different altitudes and
different transverse positions if the gratings are rolled with
regard to each other. The height of the first velocity selection
slit is $h_{\rm s}=150\,\mu\rm m$ while the third one is
restricted to $h_{\rm d}=200\,\mu\rm m$. The distance between the
source and $G_{1}$ is $L_{0}=150\,\rm{cm}$ while the spacing
between $G_{3}$ and the detector is $L_{3}=250\,\rm{mm}$. The
grating separation amounts to $L_1=L_2=105\,\rm{mm}$. The beam
width of about 1\,mm is sufficient to illuminate nearly 4000
grating openings.} \label{RollPitchYaw}
\end{figure}
Molecules passing the gratings at different heights $y_{\rm i}$
will effectively see different transverse grating shifts $\Delta
x_{\rm i}=\alpha_{\rm i}\,y_{\rm i}$ if element $i$ is rolled by
the angle $\alpha_{\rm i}$. Although a simple phase shift will not
reduce the fringe visibility, an integration over many
height-dependent shifts does. If we neglect gravity, we can
assume the second grating to define the reference angle,
$\alpha_2=0$, and study the influence of rolling G$_1$ by
$\alpha_1$ and G$_3$ around $\alpha_3$. In the paraxial limit the
total signal $\overline{S}(x)$ behind the third grating is then
\begin{eqnarray}
\overline{S}(x)&=&\frac{1}{h_{\rm s}h_{\rm d}}
\int\limits_{-h_{\rm s}/2}^{h_{\rm s}/2} \int\limits_{-h_{\rm
d}/2}^{h_{\rm d}/2} {\rm d}y_{\rm s}\,{\rm d}y_{\rm d}\,
S(x,k_{\rm d}\Delta x_{1}+k_{\rm d}\Delta x_{3})\\
&=& S_0(1+\mathcal{V}\sin(x)\,\mathrm{sinc} \left(k_{\rm d}
h_{\rm s} \alpha_1' \right)\mathrm{sinc} \left(k_{\rm d} h_{\rm d}
\alpha_3' \right)). \label{sumroll}
\end{eqnarray}
where the transmission function $S$ is parametrized as a sine wave
with visibility $\mathcal{V}$ and height dependent phase shifts
$x_{1}$ and $x_{3}$. We denote $\alpha_1' = \alpha_1
L_2(L_1+L_2+L_3)/(L_1 L_{\rm tot})+ \alpha_3 L_3/L_{\rm tot}$,
$\alpha_3' =\alpha_1 L_0 L_2/(L_1L_{\rm tot})+ \alpha_3
(L_0+L_1+L_2)/L_{\rm tot}$ with $L_{\rm tot}=(L_0+L_1+L_2+L_3)$
and all distances as shown in Figure~\ref{RollPitchYaw}.

From Eq.~(\ref{sumroll}) we see that rolling G$_1$ and G$_3$
reduces the fringe visibility in a sinc-shaped functional
dependence. The alignment becomes increasingly important for
smaller grating constants and more extended molecular beams. In
our experiment, the first zero of the sinc-curve appears for a
roll angle of around 0.65~mrad.

In addition to shifting the phase, rolling also affects the
effective grating constant. If one of the gratings is rolled with
respect to the others its projected period increases by $d_{\rm
eff}=d/\cos(\alpha) \simeq d +d \alpha^2/2  + O(\alpha^4)$
(see Fig.~\ref{RollPitchYaw}). If we require the relative period
change not to exceed $(d_{\rm eff}-d)/d = 10^{-4}$, the roll angle
has to be kept aligned to within 10\,mrad.

\subsection{Longitudinal grating shifts}
The semiclassical picture in Fig.~\ref{RollPitchYaw}a shows that the
interference pattern is also blurred when the third mask is moved
relative to the second grating by the distance $\pm \Delta L$. For
symmetry reasons, the same is true for a movement of G1. The
contrast is severely reduced when the blur $D=d/2$ is as wide as
half a grating period. We see that this condition is met when
\begin{equation}
\frac{D}{\Delta L}= \frac{d/2}{\Delta L}= \frac{Nd}{2L},
\end{equation}
where $N$ is the number of grating slits illuminated by the
molecular beam. We thus derive the {\em length-balance criterion}
\begin{equation}
\frac{\Delta L}{L}< \frac{1}{N}. \label{langenbedingung}
\end{equation}
When 4000 lines are illuminated, as in our experiments, we have to
balance the distances to better than 25\,$\mu$m. This is already
comparable to the waist of the diffracting laser beam. This
intuitive condition is consistent with a complementary and more
rigorous treatment using Wigner functions~\cite{Nimmrichter2008a}.

\subsection{Grating pitch}
The effect of forward or backward tilting a single grating, i.e.
to add a {\em pitch}, is to introduce a height-dependent
imbalance in the grating separation. The pitch must be compatible
with the requirement of Eq.~(\ref{langenbedingung}). If the beam
height is $h$ and the forward pitch is measured by the angle
$\theta$ then the arm lengths are balanced as long as $\Delta z =
h \theta \ll L/N$. For our experiment with $h=100\,\mu\rm m$,
$L=105\,\rm{mm}$ and $N=4000$, this corresponds to 250\,mrad. This
condition can easily be met.

\subsection{Grating yaw}
To first order, the argument for grating pitch also holds for
grating yaw. However, yaw also changes the slit's effective period
as well as their effective open width. Under a yaw of angle $\phi$
the grating constant shrinks like $d_{\rm y} = \cos(\phi) d
\approx d - \phi^2 d/2$. Similarly to the roll-related period
change we derive the condition $\phi < 10\,$mrad.

Finally, we have to include the effective reduction of the open
slit width if the grating is turned: The openings shrink because
of the finite wall thickness $b$ to $a_{\rm eff} = a-b\tan\phi$. A
reduction of the open fraction in the first or third grating is
important as it tends to increase the fringe visibility while
decreasing the count rate at the same time. A variation by 10\%
in visibility is already rather clearly noticeable, leading to
the constraint: $  0.1 \ge (a_{\rm eff}-a)/a = -(b/a) \tan \phi $.
For $a=90$\,nm and $b=190$\,nm we thus find a limit of $\phi=
47\,$mrad for the maximally allowed yaw angle in our experiments.

For the second grating a different reasoning applies: The optical
grating is about 20\,$\mu$m thick along the direction of the
molecular beam, therefore we have to make sure that no molecule
crosses a sizeable fraction of the standing light wave period
transversely. Since the period is as small as 266\,nm the
condition imposes a limit on both the angle of incidence and the
divergence angle of the molecular beam:  $\phi \le 0.1 \times 266
/ 20,000 \simeq 1\,$mrad.

\section{Conclusions}
Our experiment combines the virtues of near field interferometry
with the advantages of optical manipulation. Compared to far-field
diffraction of collimated beams, near-field interferometry provides
 higher signal throughput at the expense of increased
alignment requirements.  As discussed in this paper, the
experimental challenges are non-negligible but manageable with
reasonable effort. Optical phase gratings allow us to operate with
highly polarizable molecules, which would otherwise acquire
prohibitively large van der Waals phases by interacting with the
walls of material gratings.

The Kapitza-Dirac-Talbot-Lau interferometer is very well described
using a Wigner function approach, which facilitates in particular
the inclusion of grating transformations and momentum exchange in
the interferometer. The model is in excellent accordance with the
experiments, even for particles as complex as C$_{60}$F$_{36}$ and
C$_{60}$F$_{48}$. Actually the KDTLI allowed us to observe much
improved contrast compared to earlier pure Talbot Lau experiments
with the same particles~\cite{Hackermuller2003a}.

This instrument and its theoretical description are now a good basis
to continue the quest for the ultimate mass and complexity limits of
matter-wave interferometry. In addition, it has proven to be useful
for a number of relevant measurements of molecular properties, such
as optical polarizabilities and absolute optical absorption cross
sections~ \cite{Hackermuller2007a,Gerlich2008a,Nimmrichter2008b}.
Interestingly the rich internal structure of complex molecules,
including electric dipole moments or magnetic moments, structural
properties etc. can be ignored unless they either modify the optical
polarizability or unless we introduce additional field gradients or
collisions~\cite{Hornberger2003a} which allow to couple to these
properties separately~\cite{Berninger2007a}.

\section*{Acknowledgment}
We thank the Austrian science foundation FWF for support within the
projects \emph{Z149 Wittgenstein} and the doctoral program
\emph{W1210 CoQuS}. We also acknowledge financial support by the
\emph{MIME} project within the ESF Eurocore EuroQUASAR program. K.
H. is supported by the DFG within the Emmy Noether program.\\

\providecommand{\newblock}{}

\end{document}